\documentclass{ifacconf}
\usepackage{graphicx,amsmath,amssymb,color}
\usepackage{epsfig}
\usepackage[outdir=./]{epstopdf}
\graphicspath{{../eps/}}
\DeclareGraphicsExtensions{.eps}

\newcommand{\EQ}{\begin{eqnarray}}
\newcommand{\EN}{\end{eqnarray}}
\newcommand{\EQQ}{\begin{eqnarray*}}
\newcommand{\ENN}{\end{eqnarray*}}
\newcommand{\R}{\mathbb R}

\newcommand{\bremark}
{\medskip\begin{remark}
\begin{rm}}
\newcommand{\eremark}{ \end{rm}\hfill \rule{1mm}{2mm}
\end{remark} }
\newcommand{\btheorem}{\medskip\begin{theorem} \begin{it}}
\newcommand{\etheorem}{\end{it} \hfill \rule{1mm}{2mm}
\end{theorem} }
\newcommand{\blemma}{\medskip\begin{lemma} \begin{it} }
\newcommand{\elemma}{ \end{it} \hfill\rule{1mm}{2mm}
\end{lemma} }
\newcommand{\bcorollary}{\medskip\begin{corollary} \begin{it} }
\newcommand{\ecorollary}{ \end{it} \hfill\rule{1mm}{2mm}
\end{corollary} }
\newcommand{\bdefinition}{\medskip\begin{definition} }
\newcommand{\edefinition}{ \hfill\rule{1mm}{2mm}
\end{definition} }
\newcommand{\bproposition}{\medskip\begin{proposition} }
\newcommand{\eproposition}{\hfill \rule{1mm}{2mm}
\end{proposition} }
\newcommand{\bexample}{\medskip\begin{example} \begin{rm}}
\newcommand{\eexample}{ \end{rm} \hfill\rule{1mm}{2mm}
\end{example} }

\newcommand{\basm}{\medskip\begin{assumption} \begin{rm} }
\newcommand{\easm}{ \end{rm} \hfill\rule{1mm}{2mm} \medskip
\end{assumption} }
\renewcommand{\t}{^{\mbox{\tiny\sf T}}}

\newtheorem{theorem}{\sf\bfseries Theorem}[section]
\newtheorem{lemma}{\sf\bfseries Lemma}[section]

\newtheorem{definition}{\sf\bfseries Definition}[section]
\newtheorem{remark}{\sf\bfseries Remark}[section]
\newtheorem{corollary}{\sf\bfseries Corollary}[section]
\newtheorem{proposition}{\sf\bfseries Proposition}[section]
\newtheorem{example}{\sf\bfseries Example}[section]
\newtheorem{assumption}{\sf\bfseries Assumption}

\newenvironment{Proof}{\noindent{\em Proof:\/}}{\hfill $\Box$\par}
\usepackage{natbib}        
\begin{document}
\begin{frontmatter}

\title{ Analysis of Attack via Grounding and Countermeasures in Discrete-Time Consensus Networks \thanksref{footnoteinfo}} 

\thanks[footnoteinfo]{This work was supported by the Australian Research Council through
grant DP190102859.}

\author[First]{Yamin Yan} 
\author[First]{Sonja St{\"u}dli}
\author[First]{Maria M. Seron} 
\author[First]{Richard H. Middleton}

\address[First]{School of Electrical Engineering and Computing, The University of
Newcastle, Callaghan, NSW 2308, Australia, (e-mail: yamin.yan, sonja.stuedli, maria.seron, richard.middleton@newcastle.edu.au).}

\begin{abstract}                
We investigate the disruption of discrete-time consensus problems via grounding. Loosely speaking, grounding a network occurs if the state of one agent no longer responds to inputs from other agents and/or changes its dynamics. Then, the agent becomes a leader or a so called stubborn agent. The disruption of the agent can be caused by internal faults, safety protocols or externally due to a malicious attack. 
In this paper we investigate how the grounding affects the eigenratio of expander graph families that usually exhibit good scaling properties with increasing network size. It is shown that the algebraic connectivity and eigenratio of the network will decrease due to the grounding causing the performance and scalability of the network to deteriorate, even to the point of losing consensusability. We then present countermeasures to such interruptions both in a passive and active manner. Our findings are supported by numerical simulations given within the paper. 

\end{abstract}


\end{frontmatter}

\section{Introduction}

The multi-agent consensus problem has been a popular research area over the past few decades \citep{jadbabaie2003,olfati2007consensus,knorn2016}. Among the many studies on consensus, fundamental questions such as whether the network can achieve consensus (consensusability) \citep{you2011}, how to achieve consensus and consensus on what \citep{li2010consensus}, are the major topics of interest. In addition, analysis of consensus performance such as convergence rate is of both  theoretical and practical importance \citep{olfati2007consensus,chen2015convergence}.   

Efficient distributed networked control requires desirable scaling property and better consensus performance where scalability means preservation of stability of
the entire network as the network size grows large (addition of agents). Consensus networks with bounded nodal degree usually scale poorly, demonstrating scaling fragility. One type of graph family, called expander family (or expanders), scales well with bounded nodal degree \citep{pinsker1973expander}. These graphs play an important role in designing efficient communication networks. It is known that the algebraic connectivity, the second smallest eigenvalue of the graph Laplacian, is crucial in characterizing scalability and consensus performance. The algebraic connectivity of these expander families with bounded nodal degree is bounded away from zero, thus possessing desirable scalability and consensus performance. Early studies of consensus on expanders can be seen in \cite{li2009quantized}.



Recently, \cite{emma} revealed the scale fragility of expander families towards grounding in a continuous-time setting. To the best of our knowledge, the many important properties such as scalability, convergence rate, consensusability of discrete-time consensus in expander graph families towards grounding have yet to be studied.  

Grounding means that the grounded node is no longer affected by other agents while still influencing its neighbors and by doing so the complete network. Another way of interpreting this behavior is that the grounded node acts as a leader hence turning the whole network from a leaderless architecture to a leader-following one. The terminology stems from its application in power networks where grounding a node means literally connecting the bus to ground forcing the state to be set to zero. Grounding can be caused by internal faults, safety protocols or externally due to an attack. The latter would be viewed as disruption/deception attacks by either disconnecting the input channel or changing the dynamics \citep{dibaji2019systems}.




Once a network has been grounded, its dynamics can be described by a grounded Laplacian \citep{barooah2006graph}. The study of the grounded Laplacian has recently received increasing attention. For example, \cite{pirani2014spectral,pirani2015smallest,pirani2017robustness} extensively study the spectral properties of the grounded Laplacian for undirected graphs, while \cite{xia2017analysis} considers directed grounded networks. In discrete-time settings, the factor called eigenratio, i.e., the ratio of the second smallest to the largest eigenvalue of the Laplacian, \citep{you2011} plays a significant role in characterizing consensusability of undirected graph. Here we find that, for expander graph networks, while the eigenratio of the nongrounded graph is bounded away from zero with increasing network size, this no longer holds for the grounded graph. For unstable system dynamics, this reduction of the eigenratio impacts on consensusability, which in the worst case can be lost.





The contributions of this paper are three-fold. Firstly, we summarize different ways of grounding a node to turn it into a leader. Secondly, we investigate the various network properties, scalability, consensus performance as well as consensusability, of expanders towards grounding. The fragility of grounding is revealed by showing that the grounded eigenratio will decrease in network size. Thirdly, we propose countermeasures to mitigate the undesirable fragility over grounding in both a passive and active manner.



The remainder of this paper is organized as follows. We conclude this section with network graph notation and definitions. In Section~2, we provide background information on discrete-time learderless (nongrounded) and leader-following (grounded) consensus problems with a summary of network properties that this paper analyzes. In Section~3, we introduce expander families and discuss its scaling fragility towards grounding. We then show the deterioration and even loss of consensusability caused by grounding in Section~4 and possible countermeasures to mitigate the undesirable influence of grounding in Section~5. To illustrate our analysis on grounding, three numerical simulations are given in Section~6. Finally, the paper is closed with some concluding remarks in Section~7. 

\subsection*{Network graph definitions}
A graph can be described as $\mathcal{{G}}=(\mathcal{{V}},\mathcal{{E}})$, where $\mathcal{{V}}=\{1,\cdots,N\}$ is the node set and $\mathcal{{E}}$ is the edge set. A graph is undirected if the edge set consists of unordered pairs $(i,j) \in \mathcal{E}, i,j=1,\cdots,N$, if there is communication between node $i$ and node $j$, i.e. $(i,j)=(j,i)$. A graph is called simple if there are no loops ($(i,i) \not\in \mathcal{E}~ \forall i \in \mathcal{V}$) and each edge is present only once in $\mathcal{V}$. A graph family $\{\mathcal{G}_N\}$ is a sequence of graphs with increasing number of nodes, $N \rightarrow \infty$. 

If $(i,j) \in \mathcal{{E}}$, node $j$ is called a neighbor of node $i$. We use ${\mathcal{{N}}}_i = \left\{j \in \mathcal{V} \vert (i,j) \in \mathcal{E} \right\}$ to denote the neighbor set of node $i \in \mathcal{V}$. The degree of the node $i$ state the number of edges in the graph that have node $i$ as start or end. For a simple graph the degree is equal to the number of neighbors of node $i$. A graph is called $d$-regular if the degree of each node is equal to $d$. 


Let $\mathcal A=[\alpha_{ij}]_{i,j=1}^N\in\R^{N\times N}$ be the adjacency matrix of ${\mathcal{G}}$ with $i,j = 1, \cdots, N$,  $\alpha_{ii}=0$, and  $\alpha_{ij}>0$ $\Leftrightarrow$ $(j,i)\in \mathcal{ E}$. The Laplacian matrix $L=[l_{ij}]\in\R^{N\times N}$ associated with $\mathcal{G}$ is defined as $l_{ii}=\sum_{j=1}^N\alpha_{ij}$ and $l_{ij}=-\alpha_{ij}, i\neq j$.  The second smallest eigenvalue of the Laplacian matrix is called the algebraic connectivity of the graph.

\section{Preliminaries and Problem Formulation}
In this section, we first present the type of networks we consider and provide background information for the leaderless (nongrounded) and leader-following (grounded) consensus problems. We then summarize the properties and results that this paper analyzes and establishes for these problems. 

\subsection{General (leaderless) discrete-time consensus network}
We consider a discrete time multi-agent system where $N$ agents communicate among each other to achieve consensus on their states. We denote the set of agents that communicate with agent $i$ as $\mathcal{N}_i$.

Each agent is goverened by a discrete-time dynamic system given in the form of
\EQ\label{sys1}
x_i(k+1)=Ax_i(k)+Bu_i(k), k\in \mathbb{Z}^+, ~i=1,\dots,N
\EN
where $x_i\in\R^n$, $u_i\in \R$ denote the system state and control input of the $i^{th}$ agent, $A\in \R^{n\times n}$, $B\in \R^{n\times 1}$. $\mathbb{Z}^+$ denotes the set of nonnegative integers $\mathbb{Z}^+ =  \{0, 1, \cdots \}$. The standard consensus algorithm is adopted as follows,
\EQ\label{con}
u_i(k)=K\sum_{j\in\mathcal{N}_i}\alpha_{ij}(x_j(k)-x_i(k))
\EN
where $K\in \R^{1\times n}$ is the control gain matrix. 


Throughout this paper, we consider undirected, simple and connected communication graphs. This means that the adjacency matrix and the Laplacian matrix are symmetric. The adjacency matrix of the communication graph associated with~\eqref{sys1}, \eqref{con} is given as $\mathcal{A}$ with entries $\alpha_{ij} =1$ if $j \in \mathcal{N}_i$ and $\alpha_{ii} = 0$ for all $i$. $L$ is the Laplacian matrix of the communication system. The eigenvalues of $L$ are denoted by $\lambda_{i} \in \R,\ i=1,\cdots,N$ and in an ascending order are written as $0 = \lambda_{1} \leq \lambda_{2} \leq \cdots \leq \lambda_{N}$. As in \cite{you2011}, the ratio of the second smallest to the largest eigenvalue of $L$, $\lambda_{2}/\lambda_{N}$, is called the eigenratio of an undirected graph.


 The closed-loop system composed of \eqref{sys1} and \eqref{con} can be put into the following compact form
 \EQ\label{cl}
x(k+1)=(I_N\otimes A-L\otimes BK)x(k)
 \EN
where $x=[x_1^T,\dots,x_N^T]^T$ and $\otimes$ denotes the Kronecker product.

Throughout this paper, we consider the following assumptions.
\basm\label{asm2}
For $i=1,\cdots,N$, the pair $(A, B)$ is controllable and 
\EQ \label{Auinequal}
\prod_j|\lambda_j^u(A)|<\zeta^{-1}<\frac{1+\lambda_2/\lambda_N}{1-\lambda_2/\lambda_N}
\EN
for a constant $0<\zeta<1$,  where $\lambda_j^u(A)$ is an unstable eigenvalue of $A$ and the product in (4) is over all such eigenvalues.  If $A$ is stable, then $\zeta = 1$.
\easm

\basm\label{asm3}
Each node has at most $d$ neighbors, i.e., $|\mathcal{N}_i|\leq d$.
\easm

It is known that a necessary and sufficient condition for consensus of system~\eqref{cl} is that there exists $K$ such that $A-\lambda_iBK$ is Schur (i.e. all its eigenvalues are inside the open unit circle) for $i=2,\dots,N$ \citep{you2011}.
 Under Assumption~\ref{asm2}, by designing 
 \EQ\label{K}
 K=\frac{2}{\lambda_2+\lambda_N}\frac{B^TPA}{B^TPB}
 \EN where $P=P^T>0$ is a solution to the modified algebraic Riccati inequality
\EQ\label{MARI}
P-A^TPA+(1-\zeta^2)\frac{A^TPBB^TPA}{B^TPB}>0,
\EN  $A-\lambda_iBK$ will be Schur matrices for $i=2,\dots,N$. Then, consensus can be achieved with all the states $x_i(k)$ approaching $x^*=(1/N)\sum_{i=1}^Nx_i$.

\subsection{Leader-following consensus in grounded networks}

Grounding a node of a network turns the node into one that influences other nodes but is not affected in return. In a multi-agent context, this grounded node acts as a leader and converts the whole network from a leaderless architecture to a leader-following one. In other contexts, the grounded node can be interpreted as a ``stubborn agent''\citep{ghaderi2013opinion}. As mentioned previously, the terminology stems from its application in power networks where grounding a node means literally connecting the bus to ground forcing the state, in this case the voltage, to be set to $0$. To put this concept in a general framework using networked control language, we consider three different ways to ground a node, say node 1. These different ways will have different influences on network consensus.


A first form of grounding consists in fixing node 1's state $x_1(k)$ at some time $k_0$ to be either its current state or any constant value $\bar c$ in the proper dimension. Then, $x_1(k)=\bar c$ for all $k\geq k_0$. The closed-loop
system for the remaining nodes can be described as
\EQ\label{sysgrounded1}
\bar x(k+1)=&&(I_{N-1}\otimes A-\bar{L}\otimes BK)\bar x(k)\nonumber\\&&+(\Lambda\otimes BK)(\mathbf{1}_{N-1}\otimes \bar c)
\EN
where $\bar L$ is the grounded Laplacian \citep{barooah2006graph} obtained by deleting the
first row and column of $L$, $\Lambda$ is a diagonal matrix with diagonal entries equal to $\alpha_{i1}$ and $\bar x$ is obtained from $x$ by removing
the states at node 1. If $(I_{N-1}\otimes A-\bar{L}\otimes BK)$ is Schur (which will be discussed in more detail in Section~\ref{Sec4}), then $\bar x$ will approach $(I_{N-1}-(I_{N-1}\otimes A-\bar{L}\otimes BK))^{-1}(\Lambda\otimes BK)(1_{N-1}\otimes \bar c)$. 

A second form of grounding consists in cutting the control channel so that $u_1=0$, and optionally change the dynamics of node 1. Then, the first node's dynamics are $x_1(k+1)=\bar A x_1(k)$. If $\bar A=A$, the consensus trajectory will be the same as that of $x_1$ if grounding happens after the consensus was achieved; the consensus trajectory will be different from that of $x_1$ if grounding happens before the consensus achieved. 

A third form of grounding consists in taking control of $u_1$ such that $x_1$ will be steered to a deliberately designed trajectory like a certain setpoint $c_0$. Specifically, a stabilizing controller $u_1=-K_1x_1+c_1$ makes the closed-loop dynamics of node 1 $x_1(k+1)=(A-BK_1)x_1(k)+Bc_1$ with $(A-BK_1)$ Schur. The closed-loop
system for the remaining nodes will be 
\EQ\label{sysgrounded3}
\bar x(k+1)=&&(I_{N-1}\otimes A-\bar{L}\otimes BK)\bar x(k)\nonumber\\&&+(\Lambda\otimes BK)(\mathbf{1}_{N-1}\otimes x_1(k))
\EN
If $(I_{N-1}\otimes A-\bar{L}\otimes BK)$ is Schur, all states of the remaining nodes will approach $ c_0=(I-(A-BK_1))^{-1}Bc_1$.  

Note that to analyze consequences of grounding, a key system matrix to be analyzed is $(I_{N-1}\otimes A-\bar{L}\otimes BK)$. Actually, when letting $x_1=\mathbf{0}$, the closed-loop
system for the remaining nodes will be
\EQ\label{sysgrounded}
\bar x(k+1)=(I_{N-1}\otimes A-\bar{L}\otimes BK)\bar x(k).
\EN
The performance of the grounded network will be directly related to the grounded Laplacian $\bar L$. We denote the eigenvalues of $\bar L$ as $\bar \lambda_i$ and they are numbered as $0<\bar\lambda_1\leq\dots\leq\bar \lambda_{N-1}$.
The smallest eigenvalue $\bar \lambda_1$ is known as grounded eigenvalue (\cite{emma}). We denote the ratio $\bar \lambda_1/\bar \lambda_{N-1}$ by grounded eigenratio.

\subsection{Problem formulation}
In the following sections, we present results and discussions on spectral properties of $\bar L$ in regard to scalability, consensus performance, and consensusability.  Specifically, we address the following problems. 
%
%
%

\emph{Scalability of consensus:} Suppose the network graph has good scalability properties, that is, it is possible to achieve consensus as the network size grows large (addition of agents). Is scalability preserved when a node is grounded.

\emph{Consensus performance:} Is the convergence rate to consensus affected by grounding.

\emph{Consensusability for unstable systems:} How is the consensusability condition of Assumption 1 affected by grounding and can consensus be lost after grounding.

In addition, when interpreting different forms of grounding as types of attack, what are the possible countermeasures that can be taken to correct or minimise the effect of grounding.

\section{Expanders and Scalability over grounded networks}

In the following, we will investigate how the grounded network in~\eqref{sysgrounded} compares to the nongrounded network in~\eqref{cl} when using expander families as communication networks in regard to the above three properties.

For consensus networks, poor scaling property is usually observed when the network size grows large with bounded
nodal degree. An exception is when the network graph belongs to the expander family, which scale well for growing networks with bounded degree. In this section we first review the concepts of connectivity and the expander family with its desirable scaling property and then present its scaling fragility and performance degradation over grounded networks in the discrete-time setting.

 \subsection{Connectivity measures and the expander families} 
In relation to the consensus problem the algebraic connectivity of a graph, the second smallest eigenvalue of its Laplacian matrix $\lambda_{2}$, is of great interest. It is well known that the algebraic connectivity is larger than $0$ if and only if the graph is connected,  hence we find $\lambda_{2}>0$. As a measure of connectivity the algebraic connectivity is closely related to the isoperimetric constant defined below. 
\bdefinition
\citep{krebs2011expander} The \textit{isoperimetric constant} or \textit{Cheeger constant} of a graph $\mathcal{G}$ with vertex set $\mathcal{V}$ is defined as 
\EQ
h(\mathcal{G})=\min\Big\{\frac{|\partial X|}{|X|}\Big|~~ X\subset \mathcal{V} ~\mbox{and}~ |X|\leq \frac{|\mathcal{V}|}{2}\Big\}
\EN
where the boundary $\partial X$ of $X$ is the set of edges with one vertex in $X$ and the other in $\mathcal{V}-X$.
\edefinition

The isoperimetric constant is another measure of connectivity and robustness that captures how many edges need to be removed from a graph to disconnect a somewhat large number of nodes from the rest of the graph. A small number indicates the presence of a bottleneck, which is a subset of nodes connected by only few edges to the remainder of the graph. In turn this indicates a low algebraic connectivity, in fact the Cheeger inequality states \citep{lountzi2015expanders} 
\begin{equation}
  \label{eq:cheeger-inequality}
 \frac{h(\mathcal{G})^{2}}{2d} \leq \lambda_{2} \leq 2 h(\mathcal{G}).
\end{equation}

Similarly, the isoperimetric constant allows us to find a lower bound on the eigenratio of the graph
\begin{equation}
  \label{eq:eigenratio-bound}
  \frac{\lambda_{2}}{\lambda_{N}} \geq  \frac{h(\mathcal{G})^{2}}{4d^{2}}.
\end{equation}


Further, we define
\bdefinition
\citep{davidoff2003elementary} Let $c$ be a positive constant. A $d$-regular graph $\mathcal{G}$ is called a $c$-expander, if $h(\mathcal{G})\geq c$.
\edefinition

 The above concept becomes important when looking at a graph family $\{\mathcal{G}_N\}$ with $N \to \infty$. Generally, as the number of nodes increases while keeping a constant nodal degree the isoperimetric constant and the algebraic connectivity tend towards zero, which means that with increasing $N$ the graph loses its connectivity and the performance of the consensus algorithm deteriorates. An expander graph family does not exhibit this decrease in connectivity while maintaining a bounded nodal degree.
\bdefinition
\citep{krebs2011expander} (Expander family) Let $\{\mathcal{G}_N\}$ be a graph family in which $N\rightarrow \infty$. If the sequence $h(\mathcal{G}_N)$ is bounded away from $0$, $\{\mathcal{G}_N\}$ is an expander family.
\edefinition

 Note that since in an expander family $h(\mathcal{G}_{N})$ is bounded away from $0$, so is the algebraic connectivity $\lambda_{2}$ and the eigenratio $\frac{\lambda_{2}}{\lambda_{N}}$ due to the Cheeger inequality.

 \subsection{Scaling fragility and performance degradation over grounded networks}

 The advantages of using expander graphs for the consensus algorithm are clear from the previous section. These advantages are not perserved when grounding the network. 

First and foremost the scalability of the grounded network is limited. While $\lambda_2$ is bounded away from $0$ due to the property of the expander family, the larger $\lambda_2$ the better the system scales. However, in the grounded network, $\bar{\lambda}_1$ approaches zero as $N$ grows. This means that the scalability is limited.

Secondly the performace (convergence rate) of the grounded network degrades. The convergence rate directly depends on the algebraic connectivity \citep{olfati2004consensus,olfati2007consensus}. A lower algebraic connectivity indicates a slower convergence of the consensus algorithm. The following result presents the performance degradation by showing $\lambda_2 > \bar{\lambda}_1$ for large enough $N$.

%
%
\blemma \label{lemma1}
Consider a Laplacian matrix $L$ and its grounded Laplacian matrix $\bar L$, of an undirected connected regular expander graph family $\{\mathcal{G}_N\}$, then, there exists a network size $\bar N$ such that $\lambda_2>\bar \lambda_1$, for $N>\bar N$.
\elemma
\begin{Proof}
 By the eigenvalue interlacing theorem \citep{haemers1995interlacing}, we have $\lambda_2\geq \bar \lambda_1$. Now our purpose is to prove that $\lambda_2$ is strictly greater than $\bar\lambda_1$ for expander families $\{\mathcal{G}_N\}$. Suppose the graph family $\mathcal{G}_N$ is $c$-expander. By Corollary 2.3 of \cite{berman2000lower}, we have 
\EQ
\lambda_2\geq d-\sqrt{d^2-c^2}
\EN
which does not relate to the network size $N$.

If the graph is grounded, from \cite{emma}, 
\EQ
\bar \lambda_1\leq \frac{d} {N-1},
\EN
Then, $\bar \lambda_1\to 0$ for $N\to \infty$. Therefore, there exists a network size $\bar N$ such that $\lambda_2>\bar \lambda_1$ for $N>\bar N$.
\end{Proof}

\bremark
This result does not hold for all kinds of graphs. If it is not expanders, $\lambda_2$ may be equal to $\bar\lambda_1$, see the following counter example. The network contains four nodes with the following Laplacian 
\EQQ
L=\left[\begin{array}{cccc}3&-1&-1&-1\\-1&2&0&-1\\-1&0&1&0\\-1&-1&0&2 \end{array}\right]
\ENN
The eigenvalues of $L$ are $\{0,1,3,4\}$, and those of $\bar L$ by grounding the first node are $\sigma(\bar L)=\{1,1,3\}$. Clearly, $\lambda_2=\bar \lambda_1$.

\eremark

\section{Loss of Consensusability over Grounded Networks}\label{Sec4}
 
It is known that the eigenratio $\frac{\lambda_2}{\lambda_N}$ is an important factor in discrete-time networks. A larger eigenratio corresponds to better consensusability of the communication graph. Grounding causes degradation of consensus and can even disrupt consensusability. We investigate in this section when this occurs. Our argument is based on the observation that the eigenratio of the grounded network is smaller than that of the nongrounded network for large $N$. We then have the following result.

\blemma \label{lemma2}


Consider a Laplacian matrix $L$ and its grounded Laplacian matrix $\bar L$, of an undirected connected regular expander graph family $\{\mathcal{G}_N\}$, then, there exists a network size $\bar N$ such that  $\frac{\lambda_2}{\lambda_N}>\frac{\bar \lambda_1}{\bar \lambda_{N-1}}$, for $N>\bar N$.
\elemma
\begin{Proof}
By Theorem 3 of \cite{pirani2017robustness}, we have 
\EQ 
\bar \lambda_{N-1}\geq d_{\max}^{\mathcal{F}}
\EN where $d_{\max}^{\mathcal{F}}$ is the maximum degree over the nongrounded agents. 
By \cite{anderson1985eigenvalues}, 
\EQ
\lambda_N\leq \max\{d_i+d_j,~(i,j)\in\mathcal{E}\}
\EN
Then $\frac{\bar \lambda_{N-1}}{\lambda_N}\geq \frac{d_{\max}^{\mathcal{F}}}{\max\{d_i+d_j,~(i,j)\in\mathcal{E}\}}$. In particular, for a $d$-regular graph, $d_{\max}^{\mathcal{F}}=d$, $\max\{d_i+d_j,~(i,j)\in\mathcal{E}\}=2d$, then $\frac{\bar \lambda_{N-1}}{\lambda_N}\geq \frac{1}{2}$ for some $N \geq \bar N$, since $\bar \lambda_1/\lambda_2 \to 0$ for $N \to \infty$. Thus $\frac{\lambda_2}{\lambda_N}>\frac{\bar \lambda_1}{\bar \lambda_{N-1}}$, for $N>\bar N$. 



\end{Proof}

As seen in Assumption~\ref{asm2}, the eigenratio characterizes the upper bound of allowable unstable margin for discrete- time system dynamics. Grounding a network leads to a smaller eigenratio, then less unstable system dynamics will be allowed. In the case that the unstable system dynamics exceed the consensusability upper bound after grounding, the consensusability of the whole network is lost.

\begin{figure}[ht!]
\begin{center}
\includegraphics[width=8.4cm]{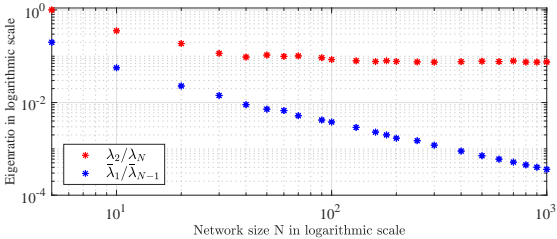}    
\caption{Eigenratio versus network size} 
\label{fig:points}
\end{center}
\end{figure}

Fig.~\ref{fig:points}  shows the nongrounded and grounded eigenratios as a function of the network size for an example of expander graph of nodal degre $4$. It can be seen that the nongrounded eigenratio is greater than the grounded eigenratio for all $N$. 

Next, a condition is given for achieving consensus for a grounded network using a controller designed for the nongrounded network. 
\blemma
 Under Assumption~\ref{asm2}, the grounded network~\eqref{sysgrounded} can achieve consensus using the nongrounded controller~\eqref{K} if all the eigenvalues $\bar L$ satisfy
\EQ\label{condition1}
&&\frac{(1-\zeta)(\lambda_2+\lambda_N)}{2}\leq\bar \lambda_i\leq\frac{(1+\zeta)(\lambda_2+\lambda_N)}{2}
\EN
for $i=1,\dots,N-1$.
\elemma 
\begin{Proof}
 To prove that $K$ in \eqref{K} also stabilizes the matrix $(I_{N-1}\otimes A-\bar L\otimes BK)$, it is equivalent to prove $A-\bar\lambda_iBK$ Schur for $i=1,\dots,N-1$. 

Direct calculation gives
\begin{align}\label{lambdai_P}
&(A -\bar\lambda_i B K)\t P (A -\bar\lambda_i B K) -P \nonumber \\
=&-P + A\t  PA   \nonumber\\
& - \left(\frac{4\bar\lambda_i}{\lambda_2+\lambda_N} -
\frac{4 \bar\lambda_i^2}{(\lambda_2+\lambda_N)^2} \right)
  \frac{A\t  P B B\t PA }{B\t PB}. 
\end{align}
If \eqref{condition1} holds, then
\EQQ
\frac{4(\lambda_2+\lambda_N)\bar\lambda_i-4\bar\lambda_i^2}{(\lambda_2+\lambda_N)^2}\geq1-\zeta^2,
\ENN
and using \eqref{MARI}, $$-P + A\t  PA - \left(\frac{4\bar\lambda_i}{\lambda_2+\lambda_N} -
\frac{4 \bar\lambda_i^2}{(\lambda_2+\lambda_N)^2} \right)
  \frac{A\t  P B B\t PA }{B\t PB}<0$$ Thus, it is proved that $A -\bar\lambda_i B K,\; i=2,\cdots, N$
is Schur, and so is the matrix $(I_{N-1}\otimes A-\bar L\otimes BK)$. The proof is completed.
\end{Proof}

\section{Discussion on grounding attacks and possible countermeasures}
As mentioned earlier, grounding may be caused by a malicious attack. For the grounding forms of fixing states or cutting input channels, the effect can be viewed as disruption attacks such as denial of service (DoS). For the grounding forms of changing the grounded node's dynamics (either by taking control or switching dynamics), the effect can be viewed as deception attacks such as false data injection (FDI) \citep{dibaji2019systems}. The latter can cause more severe effect to the whole network than the former \citep{dadras2015vehicular}.

In the previous sections we analyzed the undesirable impacts that grounding has on the network consensus, in particular, on scalability, consensus performance as well as consensusability relating to algebraic connectivity and eigenratio. In this section, we propose both passive and active countermeasures to recover from the effect of grounding.

(1)  \emph{Passive countermeasure:} design the controller beforehand to be resilient to grounding. At the stage of controller design, we select $K$ such that $(I_{N-1}\otimes A-\bar L\otimes BK)$ is Schur for every $\bar L$ resulting from grounding the $i^{th}$ node, for $i=1,\dots,N$. Then, check that $K$ also stabilizes the whole network, i.e., that $(I_N\otimes A-L\otimes BK)$ is Schur. Such a $K$ is expected to exist by noting $\bar \lambda_1<\lambda_2\leq\cdots \leq \bar \lambda_{N-1}\leq \lambda_N$, the interlacing relationship of $\lambda_i$ and $\bar \lambda_i$.
This technique may be only practical for networks of small size.

(2) \emph{Active countermeasures:} It is possible to take the following actions:

a. Suppose the grounded network is consensusable with respect to the agent dynamics. Then, redesign the controller after grounding,  that is, redesign $K$ such that $(I_{N-1}\otimes A-\bar L\otimes BK)$  is Schur. Then, consensus will be achieved.

b. Suppose the grounded network is unconsensusable with respect to the agent dynamics. If the system dynamics are unstable, and consensusability is lost, then there does not exist a $K$ to stabilize $(I_{N-1}\otimes A-\bar L\otimes BK)$. Neither redesign nor predesign the controller for grounding will work in this case. We then propose a possible approach to regain the consensusability by deliberately grounding more nodes to increase the upper bounds for the allowable unstable dynamics. This may sound counter-intuitive, but it will be seen from the following Lemma~\ref{lemma3} that by grounding more nodes, the grounded eigenvalue increases (or does not decrease), and the spectral radius of the grounded Laplacian decreases (or does not increase), thus leading to a potentially increased eigenratio and larger allowable region for the dynamics to be unstable.

\blemma \label{lemma3}




Let $\bar L^{(m)}$ be obtained by removing the first $m$ rows and $m$ columns of $L$, $m\in\mathcal{Z}^+$, $0<m<N$, $\bar \lambda_1^{(m)}$ be the smallest eigenvalue of $\bar L^{(m)}$, $\bar \lambda_{N-m}^{(m)}$ the largest eigenvalue of $\bar L^{(m)}$. Let $q \in Z^+$ be such that $0<m<q<N$. Then, $\bar \lambda_1^{(m)}\leq\bar \lambda_1^{(q)}$, $\bar \lambda_{N-m}^{(m)}\geq\bar \lambda_{N-q}^{(q)}$.
\elemma
\begin{Proof}
By the Rayleigh quotient, we have \EQ
\lambda_{\min}(\bar L^{(m)})&=&\displaystyle \min_{x_1=\dots=x_m=0}\frac{x^TLx}{x^Tx}\nonumber\\
&\leq& \displaystyle \min_{x_1=\dots=x_q=0}\frac{x^TLx}{x^Tx}=\lambda_{\min}(\bar L^{(q)})\\
\lambda_{\max}(\bar L^{(m)})&=&\displaystyle \max_{x_1=\dots=x_m=0}\frac{x^TLx}{x^Tx}\nonumber\\&\geq& \displaystyle \max_{x_1=\dots=x_q=0}\frac{x^TLx}{x^Tx}=\lambda_{\max}(\bar L^{(q)})
\EN  Thus the proof is completed.
\end{Proof}
Simulation studies suggest that the strict inequalities generally hold, that is, $\bar \lambda_1^{(m)} < \bar \lambda_1^{(q)}$, $\bar \lambda_{N-m}^{(m)}>\bar\lambda_{N-q}^{(q)}$. 

It is always possible to recover consensusability by grounding a sufficiently large number of additional  nodes for regular expanders. To see this, note that Assumption~\ref{asm2} restricts the system dynamics to be $\prod_j|\lambda_j^u(A)|<\frac{1+\lambda_2/\lambda_N}{1-\lambda_2/\lambda_N}$. If there exists $\frac{1+\bar \lambda_1^{(m)}/\bar\lambda_{N-m}^{(m)}}{1-\bar \lambda_1^{(m)}/\bar\lambda_{N-m}^{(m)}}>\prod_j|\lambda_j^u(A)|$, the network will be consensusable again.

Consider expanders with $d\geq 3$, when altogether we ground $N-2$ nodes in an extreme case, $\frac{\bar \lambda^{(m)}}{\bar \lambda_N^{(m)}}\geq\frac{1}{2}>\frac{\lambda_2}{\lambda_N}$. Then, $\prod_j|\lambda_j^u(A)|<\frac{1+\bar\lambda_1^{(m)}/\bar\lambda_{N-m}^{(m)}}{1-\bar\lambda_1^{(m)}/\bar\lambda_{N-m}^{(m)}}$ which implies that the consensusability is recovered. 

Note that the above considers a worst case estimation and in our simulations grounding only a few additional nodes $(<5)$ recovered consensusability. How many nodes to ground for regaining consensusability will be based on how unstable the system dynamics are compared to the upper bound in terms of the eigenratio. The closer $\prod_j|\lambda_j^u(A)|$ is to $\frac{1+\lambda_2/\lambda_N}{1-\lambda_2/\lambda_N}$, the more nodes will be needed to ground. It is possible that proper selection of the nodes to be grounded can reduce the necessary number. Detailed analysis of these matters is part of our future work.



\section{Numerical Simulations} 


%
\subsection{Lack of scalability over grounding}\label{Sec6.1}
Consider a leaderless vehicle platoon where the dynamics of each vehicle is modeled as a discrete-time double integrator, 
\EQ\label{vehicle}
x_{i1}(k+1)&=&x_{i1}(k)+x_{i2}(k)\nonumber\\
x_{i2}(k+1)&=&x_{i2}(k)+u_i(k),~i=1,\dots,N
\EN
where in~\eqref{vehicle}, $x_{i1}$ denotes the position from a desired setpoint, $x_{i2}$ the velocity, $u_i$ the control input of the $i^{th}$ vehicle. The system takes the form of \eqref{sys1} with $A=\left[\begin{array}{cc}1&1\\0&1 \end{array} \right]$, $B=\left[\begin{array}{cc}0\\1 \end{array} \right]$. The cooperative control objective is to have the string of vehicles travel while maintaining a certain formation, e.g., a constant target inter vehicle spacing in this example. The reference trajectory can be described as $x^*_1(k)=x^*_2\cdot k+\delta_i(k)$ with a constant speed $x^*_2=1$ and constant spacing $\delta_i(k)=5$.

\begin{figure}[ht!]
\begin{center}
\includegraphics[width=8.4cm]{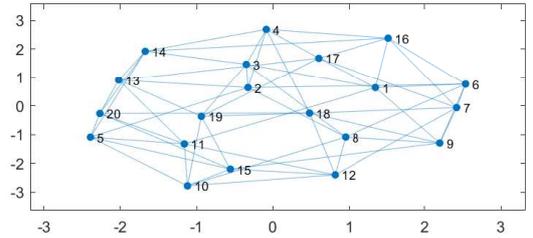}    
\caption{Random regular graph generated using the algorithm in \cite{kim2003generating} with $N=20$, $d=6$.} 
\label{fig:graph_regular}
\end{center}
\end{figure}

The communication graph is generated randomly using the algorithm in \cite{kim2003generating} with two cases, $N=20$ and $N=100$, both with degree $d=6$. See for example in Fig.~\ref{fig:graph_regular} with $20$ nodes.

\begin{figure}[ht!]
\begin{center}
\includegraphics[width=\hsize]{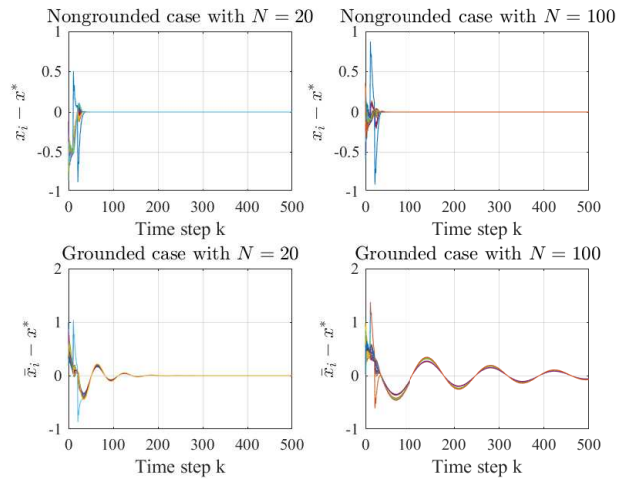}    
\caption{Profiles of deviation velocity states in nongrounded (leaderless) and grounded case (leader-following) with small and large network, perturbed by a sudden acceleration of one vehicle.} 
\label{fig:x20_100}
\end{center}
\end{figure}


The control gain is chosen as $K=[0.0157~ 0.1826]$. The results of the simulation are shown in Figs.~\ref{fig:x20_100} comparing nongrounded case (leaderless) and grounded case (leader-following) when the system has an independent lead vehicle (with dynamics $x ^*(k+1)=(A-BK)x^*(k)+Bc_1$, $c_1=1$) for two network size $N=20$ and $N=100$. During the time steps $10-20$, one of the vehicles accelerates at a doubled speed. This disturbance is attenuated by the network within a short period of time for nongrounded network. Similar performance can be observed. In contrast, for grounded networks, the disturbance is attenuated with long settling time for the smaller network and even longer for the larger one. This verifies the scalability limitation over grounded networks.

\subsection{Loss of consensusability over grounding}\label{Sec6.2}
\begin{figure}[ht!]
\begin{center}
\includegraphics[width=8.4cm]{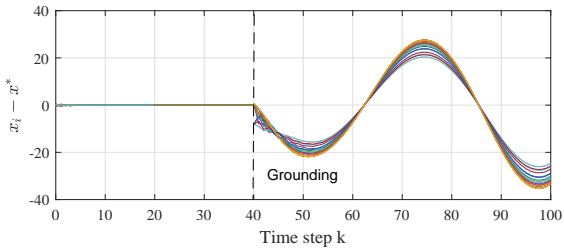}    
\caption{Profiles of deviation states: loss of consensusability after grounding } 
\label{fig:grounding}
\end{center}
\end{figure}

Consider the consensus network~\eqref{sys1} with unstable dynamics $A=\left[\begin{array}{cc}1.07&1\\0&1 \end{array} \right]$, input matrix $B=\left[\begin{array}{cc}0\\1 \end{array} \right]$. The communication graph is assumed to be the same as in Section~\ref{Sec6.1} with $N=20$. By properly designing the controller, the consensus is achieved before $k=40$. At $k=40$, one of the agents is grounded, as illustrated in Fig.~\ref{fig:grounding}. Then, the network becomes unconsensusable since the consensusability condition~\eqref{Auinequal} is not satisfied after grounding. More specifically, $\bar \delta_A=\frac{1+\bar\lambda_1/\bar\lambda_{N-1}}{1-\bar\lambda_1/\bar\lambda_{N-1}}=1.0596<\prod_{j}\left|\lambda_{j}^{u}(A)\right|=1.07< \delta_A=\frac{1+\lambda_2/\lambda_N}{1-\lambda_2/\lambda_N}=1.6935$, which means that the grounded case allows a less unstable $A$.


\subsection{Countermeasure through grounding more nodes}
Consider the same unstable network subject to the same communication graph as in Section~\ref{Sec6.2}. At $k=50$, node 1 is grounded, then the same as in Section~\ref{Sec6.2}, the consensusability is lost and the system states diverge. At $k=150$, we deliberately ground one more properly chosen node, node 2. Then, the whole network will gradually achieve consensus again since the consensusability condition~\eqref{Auinequal} can be satisfied with $\prod_{j}\left|\lambda_{j}^{u}(A)\right|=1.07<\frac{1+\bar\lambda^{(2)}_1/\bar \lambda^{(2)}_{N-2}}{1-\bar\lambda^{(2)}_1/\bar \lambda^{(2)}_{N-2}} =1.1266$. 

In an additional test, at $k=150$, instead of grounding only one more node, we deliberately ground two more nodes, nodes 2 and 3. It is observed that the consensus is achieved faster than the previous case, since $\bar \lambda_1^{(3)}=0.7026>\bar \lambda_1^{(2)}=0.5654$.

\begin{figure}[ht!]
\begin{center}
\includegraphics[width=8.4cm]{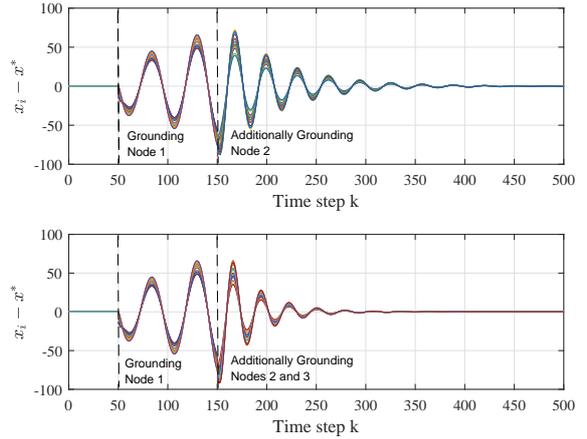}    
\caption{Above: Loss (at k=50) and regaining (after k=150) of consensusability after grounding; bottom: grounding two more nodes (Nodes 2 and 3) improves the performance (convergence rate) of consensus compared with grounding one more node} 
\label{fig:grounding_more}
\end{center}
\end{figure}

These findings are illustrated in Fig.~\ref{fig:grounding_more}.


\section{Conclusion}
In this paper, we have analyzed the scaling fragility of expanders over grounding in a discrete-time context. As in a continuous-time setting, grounded expanders do not scale well. We give a proof that the eigenratio of the grounded network will approach zero, while the one of the nongrounded expander family is bounded away from zero. This shows that the consensus performance is deteriorated and in extreme cases can even lose consensusability. We give a condition under which the grounded network is able to achieve consensus.  In addition, we discussed possible countermeasures for avoiding the loss of or regaining consensusability. The three methods discussed are to design the inital controller such that the grounded network remains stable, redesign the controller once grounding occured, or deliberately ground additional nodes. How to detect the grounded node and then either adjust the controller or select additional nodes to ground will be investigated in future work.


\bibliography{Yaminsbib}             

\begin{thebibliography}{25}
\providecommand{\natexlab}[1]{#1}
\providecommand{\url}[1]{\texttt{#1}}
\providecommand{\urlprefix}{URL }
\expandafter\ifx\csname urlstyle\endcsname\relax
  \providecommand{\doi}[1]{doi:\discretionary{}{}{}#1}\else
  \providecommand{\doi}{doi:\discretionary{}{}{}\begingroup
  \urlstyle{rm}\Url}\fi

\bibitem[{Anderson~Jr and Morley(1985)}]{anderson1985eigenvalues}
Anderson~Jr, W.N. and Morley, T.D. (1985).
\newblock Eigenvalues of the Laplacian of a graph.
\newblock \emph{Linear and multilinear algebra}, 18(2), 141--145.

\bibitem[{Barooah and Hespanha(2006)}]{barooah2006graph}
Barooah, P. and Hespanha, J.P. (2006).
\newblock Graph effective resistance and distributed control: Spectral
  properties and applications.
\newblock In \emph{Proceedings of the 45th IEEE Conference on Decision and
  Control}, 3479--3485.

\bibitem[{Berman and Zhang(2000)}]{berman2000lower}
Berman, A. and Zhang, X.D. (2000).
\newblock Lower bounds for the eigenvalues of Laplacian matrices.
\newblock \emph{Linear Algebra and its Applications}, 316(1-3), 13--20.

\bibitem[{Chen et~al.(2015)Chen, Ho, L{\"u}, and Lin}]{chen2015convergence}
Chen, Y., Ho, D.W., L{\"u}, J., and Lin, Z. (2015).
\newblock Convergence rate for discrete-time multiagent systems with
  time-varying delays and general coupling coefficients.
\newblock \emph{IEEE transactions on neural networks and learning systems},
  27(1), 178--189.

\bibitem[{Dadras et~al.(2015)Dadras, Gerdes, and Sharma}]{dadras2015vehicular}
Dadras, S., Gerdes, R.M., and Sharma, R. (2015).
\newblock Vehicular platooning in an adversarial environment.
\newblock In \emph{Proceedings of the 10th ACM Symposium on Information,
  Computer and Communications Security}, 167--178. ACM.

\bibitem[{Davidoff et~al.(2003)Davidoff, Sarnak, and
  Valette}]{davidoff2003elementary}
Davidoff, G., Sarnak, P., and Valette, A. (2003).
\newblock \emph{Elementary number theory, group theory and Ramanujan graphs},
  volume~55.
\newblock Cambridge University Press.

\bibitem[{Dibaji et~al.(2019)Dibaji, Pirani, Flamholz, Annaswamy, Johansson,
  and Chakrabortty}]{dibaji2019systems}
Dibaji, S.M., Pirani, M., Flamholz, D.B., Annaswamy, A.M., Johansson, K.H., and
  Chakrabortty, A. (2019).
\newblock A systems and control perspective of CPS security.
\newblock \emph{Annual Reviews in Control}.

\bibitem[{Ghaderi and Srikant(2013)}]{ghaderi2013opinion}
Ghaderi, J. and Srikant, R. (2013).
\newblock Opinion dynamics in social networks: A local interaction game with
  stubborn agents.
\newblock In \emph{2013 American Control Conference}, 1982--1987. IEEE.

\bibitem[{Haemers(1995)}]{haemers1995interlacing}
Haemers, W.H. (1995).
\newblock Interlacing eigenvalues and graphs.
\newblock \emph{Linear Algebra and its applications}, 226, 593--616.

\bibitem[{Jadbabaie et~al.(2003)Jadbabaie, Lin, and Morse}]{jadbabaie2003}
Jadbabaie, A., Lin, J., and Morse, A.S. (2003).
\newblock Coordination of groups of mobile autonomous agents using nearest
  neighbor rules.
\newblock \emph{IEEE Transactions on automatic control}, 48(6), 988--1001.
\newblock \doi{10.1109/TAC.2003.812781}.

\bibitem[{Kim and Vu(2003)}]{kim2003generating}
Kim, J.H. and Vu, V.H. (2003).
\newblock Generating random regular graphs.
\newblock In \emph{Proceedings of the thirty-fifth annual ACM symposium on
  Theory of computing}, 213--222. ACM.

\bibitem[{Knorn et~al.(2016)Knorn, Chen, and Middleton}]{knorn2016}
Knorn, S., Chen, Z., and Middleton, R.H. (2016).
\newblock Overview: Collective control of multiagent systems.
\newblock \emph{IEEE Transactions on Control of Network Systems}, 3(4),
  334--347.

\bibitem[{Krebs and Shaheen(2011)}]{krebs2011expander}
Krebs, M. and Shaheen, A. (2011).
\newblock \emph{Expander families and Cayley graphs: a beginner's guide}.
\newblock Oxford University Press.

\bibitem[{Li et~al.(2009)Li, Fu, Xie, and Zhang}]{li2009quantized}
Li, T., Fu, M., Xie, L., and Zhang, J.F. (2009).
\newblock Quantized consensus over expander networks and communication energy
  minimization.
\newblock In \emph{Proceedings of the 48h IEEE Conference on Decision and
  Control (CDC) held jointly with 2009 28th Chinese Control Conference},
  5809--5814. IEEE.

\bibitem[{Li et~al.(2010)Li, Duan, Chen, and Huang}]{li2010consensus}
Li, Z., Duan, Z., Chen, G., and Huang, L. (2010).
\newblock Consensus of multiagent systems and synchronization of complex
  networks: A unified viewpoint.
\newblock \emph{IEEE Transactions on Circuits and Systems I: Regular Papers},
  57(1), 213--224.

\bibitem[{Lountzi(2015)}]{lountzi2015expanders}
Lountzi, A. (2015).
\newblock \emph{Expander Graphs and Explicit Construction}.
\newblock Ph.D. thesis, Uppsala Universitet.

\bibitem[{Olfati-Saber et~al.(2007)Olfati-Saber, Fax, and
  Murray}]{olfati2007consensus}
Olfati-Saber, R., Fax, J.A., and Murray, R.M. (2007).
\newblock Consensus and cooperation in networked multi-agent systems.
\newblock \emph{Proceedings of the IEEE}, 95(1), 215--233.

\bibitem[{Olfati-Saber and Murray(2004)}]{olfati2004consensus}
Olfati-Saber, R. and Murray, R.M. (2004).
\newblock Consensus problems in networks of agents with switching topology and
  time-delays.
\newblock \emph{IEEE Transactions on automatic control}, 49(9), 1520--1533.

\bibitem[{Pinsker(1973)}]{pinsker1973expander}
Pinsker, M.S. (1973).
\newblock On the complexity of a concentrator.
\newblock In \emph{7th International Teletraffic Conference}.

\bibitem[{Pirani et~al.(2017)Pirani, Shahrivar, Fidan, and
  Sundaram}]{pirani2017robustness}
Pirani, M., Shahrivar, E.M., Fidan, B., and Sundaram, S. (2017).
\newblock Robustness of leader--follower networked dynamical systems.
\newblock \emph{IEEE Transactions on Control of Network Systems}, 5(4),
  1752--1763.

\bibitem[{Pirani and Sundaram(2014)}]{pirani2014spectral}
Pirani, M. and Sundaram, S. (2014).
\newblock Spectral properties of the grounded Laplacian matrix with
  applications to consensus in the presence of stubborn agents.
\newblock In \emph{2014 American Control Conference}, 2160--2165. IEEE.

\bibitem[{Pirani and Sundaram(2015)}]{pirani2015smallest}
Pirani, M. and Sundaram, S. (2015).
\newblock On the smallest eigenvalue of grounded Laplacian matrices.
\newblock \emph{IEEE Transactions on Automatic Control}, 61(2), 509--514.

\bibitem[{Tegling et~al.(2019)Tegling, Middleton, and Seron}]{emma}
Tegling, E., Middleton, R.H., and Seron, M.M. (2019).
\newblock Scalability and fragility in bounded-degree consensus networks.
\newblock In \emph{8th IFAC Workshop on Distributed Estimation and Control in
  Networked Systems}.

\bibitem[{Xia and Cao(2017)}]{xia2017analysis}
Xia, W. and Cao, M. (2017).
\newblock Analysis and applications of spectral properties of grounded
  Laplacian matrices for directed networks.
\newblock \emph{Automatica}, 80, 10--16.

\bibitem[{You and Xie(2011)}]{you2011}
You, K. and Xie, L. (2011).
\newblock Network topology and communication data rate for consensusability of
  discrete-time multi-agent systems.
\newblock \emph{IEEE Transactions on Automatic Control}, 56(10), 2262--2275.

\end{thebibliography}
                                                   







\end{document}